\documentclass{article}

\begin{document}

\centerline{ON THE GRAVITATIONAL POTENTIAL OF AN INHOMOGENEOUS}
\centerline{ELLIPSOID OF REVOLUTION (SPHEROID)}

\vskip 5mm

\centerline{V.V.Gvaramadze}

\vskip 3mm

\centerline{{\it Sternberg Astronomical Institute, Moscow State
University,  Moscow,  Russia;}} \centerline{{\it
vgvaram@sai.msu.ru}}

\vskip 3mm

\centerline{and}

\vskip 3mm

\centerline{J.G.Lominadze}

\vskip 3mm

\centerline{{\it Abastumani Astrophysical Observatory, Georgian
Academy of Sciences,}} \centerline{{\it Tbilisi, Georgia}}

\vskip 5mm

\centerline{Abstract}

\vskip 5mm

It is shown that the gravitational potential outside an
inhomogeneous ellipsoid of revolution (spheroid) whose isodensity
surfaces are confocal spheroids is identical to the gravitational
potential of a homogeneous spheroid of the same mass.

\vskip 5mm

\section{Introduction}

The result presented in this note was published in our paper
entitled ``Rotation of gas above the Galactic disk" (Astrophysics,
1988,  28, 57-64; see also Proceedings of the Joint
Varenna-Abastumani International School and Workshop
on Plasma Astrophysics, 1986, ESA SP-251, 551-556).
(Below we give Sect.\,2 of this paper, as it appeared in
Astrophysics.) Recently B.P.Kondratyev (2003, The Potential Theory and
Equilibrium Figures\footnote{We thank O.G.Chkhetiani for bringing
this monography to our attention.}, Moskow-Izhevsk: The Institute
of Computer Science, 624 p.) showed that the same result is valid
in the more general case of confocal triaxial ellipsoids.

\section{The Gravitational Potential of the Disk}

In this section, we obtain the gravitational potential of an
ellipsoid of revolution having as isodensity surfaces confocal
ellipsoids. The gravitational potential of a body of arbitrary
shape is
\begin{equation}
\Phi ({\vec r}) \, = \, -G\int {dM \over R} \, = \, G\int
{\rho({\vec {r^{'}}})d^3r^{'} \over |{\vec r}-{\vec {r^{'}}}|} \,
,
\end{equation}
where the integration is over the complete volume of the body. If
the body possesses some symmetry (exact or approximate), then the
most effective method for finding the potential is an expansion
with respect to orthogonal functions. The particular choice of the
orthogonal system of functions depends on the symmetry.

We shell assume that the stellar disk of the Galaxy has the form
of an oblate ellipsoid of revolution (spheroid). The section of
the disk perpendicular to the plane of rotation has the form of an
ellipse. It is then natural to make all the calculations in a
system of oblate spheroidal coordinates:
\begin{equation}
\left\{
\begin{array}{ll}
x=c[(\xi ^2 +1)(1-\eta ^2 )]^{1/2} \, \cos \phi \, , & \\
y=c[(\xi ^2 +1)(1-\eta ^2 )]^{1/2} \, \sin \phi \, , & \\
z=c\xi \eta \, , \, 0\leq \xi < \infty , -1\leq \eta <1 \, , 0\leq
\phi \leq 2\pi \, . &
\end{array}
\right.
\end{equation}
The Lam\'e parameters in this system of coordinates have the form
$$
h_{\xi} =c\left[ {\xi ^2 + \eta ^2 \over \xi ^2 +1}\right]^{1/2} ,
\, h_{\eta} =c\left[ {\xi ^2 + \eta ^2 \over 1 - \eta
^2}\right]^{1/2} , \, h_{\phi} =c[(\xi ^2 + 1)(1- \eta ^2 )]^{1/2}
\, ,
$$
and the element of volume can be expressed in terms of the
introduced coordinates as follows:
\begin{equation}
d^3 r = c^3 (\xi ^2 + \eta ^2 ) d\xi d\eta d\phi \, ,
\end{equation}
where $c=\sqrt{a^2 -b^2}$ is the half-distance between the focuses
of the spheroid, and $a$ and $b$ are, respectively, the semimajor
and semiminor axes of the spheroid. The boundary of the spheroid
is determined by the relation $\xi = \xi _0 =b/c$.

In the oblate spheroidal coordinates we can expand $R^{-1}$ in a
series in associated Legendre functions [7,8]:
$$
{1 \over R} \, = \, {1 \over c} \, \sum _{n=0} ^{\infty} (2n+1)
\sum _{m=0} ^{n} \epsilon _m i^{m+1} \left[{(n-m)! \over
(n+m)!}\right]^2 \cos [m(\phi -\phi ^{'} )]\times
$$
\begin{equation}
P_n ^m (\eta ^{'} )P_n ^m (\eta ) \left\{
\begin{array}{ll}
P_n ^m (i\xi ^{'} ) Q_n ^m (i\xi )\, , & \xi > \xi ^{'} \, , \\
P_n ^m (i\xi ) Q_n ^m (i\xi ^{'} )\, , & \xi ^{'} > \xi \, ,
\end{array}
\right.
\end{equation}
where
$$
\left\{
\begin{array}{ll}
\epsilon _m =1, & {\rm for\ } m=0 \, ,\\
\epsilon _m =2, & {\rm for\ } m>0 \, ,
\end{array}
\right.
$$
$P_n ^m$ and $Q_n ^m$ are associated Legendre functions
of the first and second kinds, respectively.

Substituting the expansion (4) for the case $\xi > \xi _0 \geq \xi
^{'}$, i.e., outside the disk, in the general formula (1) and
taking into account (3), we obtain a representation of the
gravitational potential of the ellipsoid in the oblate spheroidal
coordinates:
$$
\Phi (\xi , \eta , \phi ) \, = \, - Gc^2 \int \rho (\xi ^{'}, \eta
^{'}, \phi ^{'} )(\xi ^{'2} + \eta ^{'2} ) \sum _{n=0} ^{\infty}
(2n+1) \sum _{m=0} ^{n} \epsilon _m i^{m+1} \times
$$
$$
\left[{(n-m)! \over (n+m)!}\right]^2 \cos [m(\phi -\phi ^{'} )] \,
P_n ^m (\eta ^{'} ) P_n ^m (i\xi ^{'}) P_n ^m (\eta ) Q_n ^m (i\xi
)d\xi ^{'} d\eta ^{'} d\phi ^{'} \, .
$$
The integration over $\xi ^{'}$ is from $0$ to $\xi _0$, over
$\eta ^{'}$ from $-1$ to $+1$, and over $\phi$ from $0$ to $2\pi$.

We assume that the density does not depend on $\phi$ (axial
symmetry), and then all integrals with $m>0$ are zero. The
expression for the gravitational potential simplifies,
$$
\Phi (\xi , \eta ) \, = \, - Gc^2 \int \rho (\xi ^{'}, \eta ^{'}
)(\xi ^{'2} + \eta ^{'2} ) \times
$$
$$
\left[i\sum _{n=0} ^{m} (2n+1)P_n (\eta ^{'} ) P_n (i\xi ^{'})P_n
(\eta ) Q_n (i\xi )\right]d\xi ^{'} d\eta ^{'} d\phi ^{'} \, .
$$
We consider the terms of this series
\begin{equation}
\Phi (\xi , \eta ) \, = \, \Phi _0 (\xi , \eta ) \, + \,\Phi _1
(\xi , \eta ) \, + \, \Phi _2 (\xi , \eta ) \, + \, ... \, ,
\end{equation}
where $\Phi _0, \Phi _1, \Phi _2$, etc., are determined by
\begin{equation}
\left\{
\begin{array}{ll}
\Phi _0 (\xi , \eta ) \, = \, -c^2 GI_0 {\rm arctg} (1/\xi ) \, , & \\
I_0 \, = \, 2\pi \int _0 ^{\xi _0} \int _{-1} ^1 \rho (\xi ^{'},
\eta ^{'} )(\xi ^{'2} + \eta ^{'2} )P_0 (\eta ^{'}) P_0 (i\xi
^{'})d\xi ^{'} d\eta ^{'} \, ; &
\end{array}
\right.
\end{equation}
$$
\left\{
\begin{array}{ll}
\Phi _1 (\xi , \eta ) \, = \, -3c^2 GI_1 \eta [\xi {\rm arctg}
(1/\xi )
-1] \, , & \\
I_1 \, = \, 2\pi i \int _0 ^{\xi _0} \int _{-1} ^1 \rho (\xi ^{'},
\eta ^{'} )(\xi ^{'2} + \eta ^{'2} )P_1 (\eta ^{'}) P_1 (i\xi
^{'})d\xi ^{'} d\eta ^{'} \, ; &
\end{array}
\right.
$$
\begin{equation}
\left\{
\begin{array}{ll}
\Phi _2 (\xi , \eta ) \, = \, {5\over 4}c^2 GI_2 (3\eta ^2 -1)
[(1+3\xi ^2 ){\rm arctg} (1/\xi ) -3\xi] \, , & \\
I_2 \, = \, 2\pi \int _0 ^{\xi _0} \int _{-1} ^1 \rho (\xi ^{'},
\eta ^{'} )(\xi ^{'2} + \eta ^{'2} )P_2 (\eta ^{'}) P_2 (i\xi
^{'})d\xi ^{'} d\eta ^{'} \, . &
\end{array}
\right.
\end{equation}
The constants $I_0 , I_1 , I_2$, etc., can be readily determined
by specifying the particular form of the density function. For
$\rho =\rho (\xi )$, i.e., when the density is distributed over
confocal spheroids (and also in the special case $\rho = {\rm
const}$), all the constants except $I_0$ and $I_2$ are zero. This
can be readily seen by representing $\xi ^2 + \eta ^2$ in the form
$$
\xi ^2 + \eta ^2 \equiv P_0 (\eta ) \left(\xi ^{2} +{1\over
3}\right) + {2\over 3} P_2 (\eta)
$$
and by using in the integration over $\eta$ the orthogonality
property of the Legendre functions:
$$
\int _{-1} ^1 P_n (\eta ) P_m (\eta )d\eta \, = \, \delta _{nm} \,
{2\over 2n+1} \, .
$$
In the well-known (see, for example, [9]) special case $\rho =
{\rm const}$, the constants $I_0$ and $I_2$ have the form
\begin{equation}
I_0 \, = \, {4\pi \over 3} \rho \xi _0 (1+\xi _0 ^2 ) \, = \,
{1\over c^3}\rho V \, = \, {M\over c^3} \, ,
\end{equation}
\begin{equation}
I_2 \, = \, -{1\over 5} I_0 \, .
\end{equation}
In (8) $V=(4\pi /3)a^2 b$ is the volume of the spheroid, and $M$
is its mass.

After substitution of (6), (7), and (9) in (5) we obtain
\begin{equation}
\Phi (\xi , \eta ) \, = \, -{GM \over c} \left\{ {\rm arctg}
(1/\xi ) + {1 \over 4}(3\eta ^2 -1)[(1+3\xi ^2) {\rm arctg} (1/\xi
) -3\xi ] \right\} \, .
\end{equation}
The oblate spheroidal coordinates $\xi$ and $\eta$ are related to
the cylindrical coordinates $r$ and $z$ by
$$
\xi \, = \, {1\over {\sqrt 2} c} \, [\kappa +\rho ]^{1/2} \, , \,
\eta \, = \, {1\over {\sqrt 2} c} \, [\kappa -\rho ]^{1/2} \, , \,
$$
$$
\kappa \, = \, [\rho ^2 + 4z^2 c^2 ]^{1/2} \, , \, \rho \, = \,
z^2 +r^2 -c^2 \, .
$$
After some manipulations the expression (10) can be shown to be
identical to the well-known expression for the gravitational
potential of a homogeneous spheroid [9]. However, it is found that
this expression is unchanged in the more general case $\rho = \rho
(\xi )$, since the relation (9) between the constants $I_0$ and
$I_2$ remains true. Thus, the gravitational potential outside an
inhomogeneous ellipsoid of revolution whose isodensity surfaces
are confocal ellipsoids is identical to the gravitational
potential of a homogeneous ellipsoid of revolution of the same
mass. Similarly, the gravitational potential outside a spherically
symmetric mass distribution does not depend on the particular
distribution of the density.

We note that the gravitational potential of an ellipsoid of
revolution for the special case of isodensity surfaces in the form
of confocal ellipsoids was obtained in [10], but the identity of
the obtained potential and the potential of a homogeneous
ellipsoid was not pointed out.

$<...>$

\vskip 5mm

We are grateful to A.A.Ruzmaikin and A.M.Shukurov for numerous
helpful discussions, to M.G.Abramyan and A.M.Fridman for their
interest in the work, and also V.L.Polyachenko, who drew our
attention to [10].

\vskip 5mm

\centerline{LITERATURE CITED}

\vskip 5mm

\parindent=0pt

$<...>$\\

7. E.W.Hobson, The Theory of Spherical and Ellipsoidal
Harmonics, Cambridge (1931)\\

8. E.M.Morse and H.Feshbach, Methods of Theoretical
Physics, Vol. 2, McGraw-Hill, New York (1953)\\

9. S.Chandrasekhar, Ellipsoidal Figures of Equilibrium, New Haven
(1969)\\

10. L.Perek, Bull. Astron. Inst. Czech., 9, 212 (1958)\\

$<...>$

\end{document}